\documentclass[conference]{IEEEtran}
\IEEEoverridecommandlockouts

\usepackage{cite}
\usepackage{amsmath,amssymb,amsfonts}
\usepackage{algorithmic}
\usepackage{graphicx}
\usepackage{textcomp}
\usepackage{xcolor}
\usepackage[misc]{ifsym}

\def\BibTeX{{\rm B\kern-.05em{\sc i\kern-.025em b}\kern-.08em
    T\kern-.1667em\lower.7ex\hbox{E}\kern-.125emX}}

\begin{document}

\title{MPE-Adam: Multi-Population Evolutionary Optimization with Adam Refinement for QAOA}

\author{\IEEEauthorblockN{Chi Quan Luu\textsuperscript{\Letter}}
\IEEEauthorblockA{\textit{University of Wollongong} \\
Wollongong, Australia \\
samluu206@gmail.com \\
ORCID: 0009-0002-3185-0617}
\and
\IEEEauthorblockN{Thai T. Vu\textsuperscript{\Letter}}
\IEEEauthorblockA{\textit{University of Wollongong} \\
Wollongong, Australia \\
tienv@uow.edu.au \\
ORCID: 0000-0002-9826-3321}
\and
\IEEEauthorblockN{John Le}
\IEEEauthorblockA{\textit{University of Wollongong} \\
Wollongong, Australia \\
johnle@uow.edu.au \\
ORCID: 0000-0003-0019-0345}
}

\maketitle

\begin{abstract}
Parameter optimization is a central bottleneck in variational quantum algorithms such as the Quantum Approximate Optimization Algorithm (QAOA). The classical optimizer must navigate a high-dimensional, non-convex parameter space under measurement noise. From a quantum software perspective, this process forms a multi-stage workflow: global exploration of the parameter space followed by local refinement within the hybrid quantum-classical loop. Most existing approaches, however, employ single-stage optimizers that do not separate these roles, which limits the use of complementary strategies. We propose MPE-Adam, a hybrid optimization framework that integrates multi-population evolutionary search for global exploration with Adam-based gradient refinement for local convergence. The method is structured as a modular component suitable for quantum software pipelines. We evaluate MPE-Adam on MaxCut instances generated from random 3-regular graphs with up to 22 nodes. The results show that MPE-Adam achieves higher approximation ratios and lower variance than evolutionary-only and SPSA-based baselines, with statistically significant improvements. These findings indicate that structured multi-stage optimization improves both solution quality and software-level flexibility in quantum applications.
\end{abstract}


\section{Introduction}

Hybrid quantum--classical methods, known as variational quantum algorithms (VQAs), have attracted significant attention in the NISQ era. The Quantum Approximate Optimization Algorithm (QAOA) \cite{farhi2014quantum} is a representative example that targets combinatorial problems such as MaxCut \cite{zhou2020quantum}. A broad overview of QAOA variants and their trade-offs is given in \cite{blekos2024review}. These algorithms rely on parameterized quantum circuits whose parameters are iteratively tuned through a hybrid quantum-classical loop \cite{mcclean2016theory}. Parameter optimization remains a major bottleneck. The objective landscape is high-dimensional and non-convex, and expectation values are estimated from finite measurements affected by hardware noise \cite{mcclean2018barren,larocca2025barren}. Guerreschi and Matsuura \cite{guerreschi2019qaoa} estimated that hundreds of qubits may be needed before QAOA achieves a genuine speed-up over classical MaxCut solvers, which underscores the need for effective parameter optimization at all scales.

From an optimization perspective, parameter tuning in VQAs is inherently multi-stage. It involves initialization, global exploration of the parameter space, and local refinement to improve solution quality. Most existing approaches rely on single-stage strategies. Derivative-free methods such as COBYLA \cite{powell1994direct} are popular defaults but often plateau on the noisy, multimodal cost surfaces typical of variational circuits. Gradient-based methods such as Adam \cite{kingma2015adam} and SPSA \cite{spall1992multivariate} provide fast convergence but are sensitive to initialization and prone to local optima \cite{mcclean2018barren}. Monta\~{n}ez-Barrera and Michielsen \cite{montanez2025linear} showed that even fixed linear-ramp schedules can yield competitive approximation ratios once enough circuit layers are used, which highlights the importance of parameter strategy choice. Evolutionary and population-based methods offer stronger global exploration but typically require many function evaluations and lack efficient local refinement \cite{schiavello2025evolving}.

Evolutionary algorithms are widely used for global search in VQAs because they are gradient-free and robust to noise \cite{schiavello2025evolving}. Multi-population evolutionary strategies further enhance diversity by maintaining several interacting subpopulations \cite{kolle2025evaluating,alvarado2025quantum}. Adaptive gradient-based optimizers such as Adam are effective at refining parameters once a promising region has been identified \cite{yang2025quantum,gacon2023empirical}. These approaches are typically applied independently or loosely combined, without a unified framework that explicitly coordinates exploration and refinement.

This gap motivates a structured multi-stage optimization strategy that integrates global exploration with efficient local convergence.

We propose MPE-Adam, a hybrid optimization framework that combines multi-population evolutionary search with Adam-based local refinement. The method separates and coordinates exploration and exploitation. Population diversity drives exploration of the parameter space, while gradient information accelerates convergence toward high-quality solutions.

We evaluate MPE-Adam on MaxCut benchmark instances generated from random 3-regular graphs with up to 22 nodes \cite{zhou2020quantum}. The results show that MPE-Adam consistently achieves higher approximation ratios and lower variance than evolutionary-only and SPSA-based baselines, with statistically significant improvements.

The main contributions of this work are:
\begin{itemize}
\item We formulate QAOA parameter optimization as a structured multi-stage process that distinguishes global exploration from local refinement.
\item We propose MPE-Adam, a hybrid framework that integrates multi-population evolutionary search with Adam-based refinement in a unified design.
\item We demonstrate improved solution quality and stability on QAOA-based MaxCut problems through comprehensive experimental evaluation.
\end{itemize}

\section{Related Work}

\textbf{QAOA extensions.} After the original proposal \cite{farhi2014quantum}, many circuit-level modifications have emerged. Hadfield et al. \cite{hadfield2019quantum} generalized the QAOA mixer to a quantum alternating operator ansatz, which allows constraints to be enforced natively within the circuit. Bravyi et al. \cite{bravyi2020obstacles} developed RQAOA, which progressively pins decision variables based on measured correlations and shows clear benefits on dense graphs. Herrman et al. \cite{herrman2022multi} introduced multi-angle QAOA, which assigns independent rotation angles to every gate and improves accuracy at modest additional parameter cost. Monta\~{n}ez-Barrera and Michielsen \cite{montanez2025linear} demonstrated that fixed linear-ramp parameter schedules can compete with classically optimized QAOA on weighted MaxCut instances up to 42 qubits. A broad overview of QAOA variants is provided in \cite{blekos2024review}.

\textbf{Gradient-based and stochastic optimizers.} The outer-loop optimizer plays a decisive role in variational quantum performance. COBYLA \cite{powell1994direct}, a derivative-free simplex method, remains widely adopted. SPSA \cite{spall1992multivariate} offers a stochastic gradient signal from only two cost queries, which makes it well suited to noisy quantum backends \cite{pellow2024effect}. Pellow-Jarman et al. \cite{pellow2024effect} benchmarked classical optimizers on QAOA under realistic noise and found that SPSA, Adam, and AMSGrad consistently outperform other methods. Gacon et al. \cite{gacon2023empirical} paired SPSA gradient signals with momentum-based rules such as Adam and AMSGrad. Their results show that Adam in particular accelerates convergence and lowers the final error for variational eigensolvers. Zhou et al. \cite{zhou2020quantum} conducted an in-depth study of QAOA on MaxCut and proposed heuristic parameter-initialization strategies, providing further evidence that adaptive optimizers become essential as ansatz complexity grows. These methods are nevertheless sensitive to initialization and may converge to suboptimal local optima in highly non-convex landscapes.

\textbf{Population-based and evolutionary methods.} Population-based heuristics offer an orthogonal strategy. Wang et al. \cite{wang2023pid} applied a quantum-inspired genetic algorithm to PID controller tuning. Schiavello et al. \cite{schiavello2025evolving} demonstrated that an island-model EA with CVaR-based fitness can rival or outperform COBYLA on QAOA-based MaxCut. The CVaR aggregation function, originally proposed by Barkoutsos et al. \cite{barkoutsos2020improving}, replaces the standard expectation value with a tail-risk metric. This focuses optimization pressure on the best measurement outcomes and leads to faster convergence and higher-quality solutions on combinatorial tasks. Shaydulin et al. \cite{shaydulin2024evidence} recently provided evidence of a scaling advantage for QAOA on a classically intractable problem using trapped-ion hardware, which highlights the importance of sample-efficient hybrid optimizers for larger instances. These approaches typically require a large number of evaluations and lack efficient local refinement.

\textbf{Hybrid optimization strategies.} Recent work has begun to explore combinations of global and local optimization. Evolutionary QAOA variants such as E-QAOA \cite{schiavello2025evolving} demonstrate the benefits of population-based exploration. Integrating global search with efficient gradient-based refinement, however, remains an open challenge. In this work, we extend this line by appending an Adam refinement stage to a multi-population evolutionary framework, enabling both effective exploration and efficient local convergence. We further provide a four-way comparison that includes SPSA and Adam local polishing on problem sizes large enough to reveal meaningful performance differences.

\section{Problem Statement}

We study the optimization of Quantum Approximate Optimization Algorithm (QAOA) parameters for solving the MaxCut problem on a graph $G = (V, E)$. The MaxCut objective can be written as
\begin{equation}
C(z) = \sum_{(i,j) \in E} \frac{1 - z_i z_j}{2},
\end{equation}
where $z_i \in \{-1, 1\}$. This problem is NP-hard; the tightest known polynomial-time guarantee is the 0.878-factor of Goemans and Williamson \cite{goemans1995improved}, obtained via semidefinite relaxation. In QAOA, this objective is encoded into a cost Hamiltonian $H_C$, and a parameterized quantum state $|\psi(\boldsymbol{\gamma}, \boldsymbol{\beta})\rangle$ is prepared using variational angles $\boldsymbol{\gamma}$ and $\boldsymbol{\beta}$ \cite{farhi2014quantum}. The goal is to maximize the expected value of $H_C$:
\begin{equation}
F(\boldsymbol{\gamma}, \boldsymbol{\beta}) = \langle \psi(\boldsymbol{\gamma}, \boldsymbol{\beta}) | H_C | \psi(\boldsymbol{\gamma}, \boldsymbol{\beta}) \rangle,
\end{equation}
i.e.,
\begin{equation}
(\boldsymbol{\gamma}^*, \boldsymbol{\beta}^*) = \arg\max_{\boldsymbol{\gamma}, \boldsymbol{\beta}} F(\boldsymbol{\gamma}, \boldsymbol{\beta}).
\end{equation}
We evaluate performance using the approximation ratio
\begin{equation}
\rho = \frac{F(\boldsymbol{\gamma}, \boldsymbol{\beta})}{C_{\max}}.
\end{equation}

This optimization problem is highly non-convex and prone to local optima. Gradient-based methods such as Adam and SPSA offer fast convergence but are sensitive to initialization. Evolutionary methods provide better global exploration but often lack efficient local refinement. The key challenge is to design a strategy that balances exploration and exploitation in the QAOA parameter space.

\section{Proposed Method}

MPE-Adam follows a two-stage optimization pipeline that combines global exploration with local refinement for QAOA parameter tuning. Stage 1 is the global exploration phase. It uses a multi-population evolutionary algorithm in which several independent populations evolve in parallel and periodically exchange elite individuals through migration every $g_f$ generations. This island-model strategy enables diverse exploration of the parameter space and identifies promising candidate solutions. Stage 2 is the local refinement phase. The best candidate from the evolutionary process is further optimized using the Adam optimizer, which leverages gradient information and momentum-based updates to improve convergence and solution quality. By combining population-based global search with adaptive gradient-based refinement, MPE-Adam balances exploration and exploitation.

\subsection{Multi-Population Evolutionary Framework}

The island model maintains $M$ independent populations evolving concurrently. Each population $P_m$ ($m = 1, \dots, M$) holds $N_{\text{pop}}$ individuals, where each individual $\theta_i = (\boldsymbol{\gamma}_i, \boldsymbol{\beta}_i)$ encodes a candidate set of QAOA angles.

At each generation, the evolutionary process within each population proceeds as follows:

\textbf{Initialization:} Angles are drawn uniformly with $\gamma_s, \beta_s \in [-\pi, \pi]$ for every QAOA layer $s$.

\textbf{Fitness Evaluation:} The QAOA circuit corresponding to each individual is executed and the approximation ratio (4) is recorded. We adopt CVaR at the $\alpha$-quantile as the fitness metric \cite{barkoutsos2020improving}, concentrating selection pressure on the distribution's upper tail.

\textbf{Selection:} Parents are chosen via Stochastic Universal Sampling (SUS), which provides uniform selection pressure across the fitness distribution.

\textbf{Crossover and Mutation:} Offspring are created by whole arithmetic crossover of selected parents, blending parental angles with a random weight $\alpha \sim U(0,1)$, followed by self-adaptive Gaussian mutation in which each individual's step-size vector $\sigma$ co-evolves via log-normal adaptation with learning rates $\tau$ and $\tau'$ set per B\"{a}ck's standard formulation \cite{back1996evolutionary} and a minimum threshold $\sigma_{\min}$.

\textbf{Elite Migration:} Every $g_f$ generations, the top $k_{\text{elite}}$ individuals from each island are broadcast to all other islands, replacing their weakest members. This mechanism balances knowledge transfer with population diversity.

In both the evolutionary and Adam refinement stages, angles are wrapped periodically into $[-\pi, \pi]$ after each update.

\subsection{Adam Local Refinement}

After the evolutionary process terminates at generation $G_{\max}$, the globally best individual $\theta^*_{\text{EA}}$ is selected and used to initialize the Adam optimizer. Gradients are estimated using central finite differences with perturbation $\varepsilon = 0.01$, which requires $2d$ circuit evaluations per Adam step for $d = 2p$ parameters. The parameters are then refined over $T_{\text{Adam}}$ iterations according to the update rules in (5)--(8).
\begin{align}
\mu_k &= \beta_1 \mu_{k-1} + (1 - \beta_1) g_k, \\
\nu_k &= \beta_2 \nu_{k-1} + (1 - \beta_2) g_k^2,
\end{align}
Since both accumulators start at zero, a bias-correction step compensates for the initial under-estimation:
\begin{equation}
\hat{\mu}_k = \frac{\mu_k}{1 - \beta_1^k}, \quad \hat{\nu}_k = \frac{\nu_k}{1 - \beta_2^k}.
\end{equation}
Each parameter is then updated via:
\begin{equation}
\theta_{k+1} = \theta_k - \eta \frac{\hat{\mu}_k}{\sqrt{\hat{\nu}_k} + \epsilon},
\end{equation}
with learning rate $\eta$ and stability constant $\epsilon = 10^{-8}$.

We set $\beta_1 = 0.9$, $\beta_2 = 0.999$, step size $\eta = 0.01$, and stability constant $\epsilon = 10^{-8}$, consistent with the standard Adam configuration \cite{kingma2015adam}. By adapting the effective step size for each parameter, Adam handles the heterogeneous curvature between $\gamma$ and $\beta$ angles without manual tuning.

\subsection{MPE-Adam Algorithm}

\begin{figure*}[t]
\centering
\includegraphics[width=\textwidth]{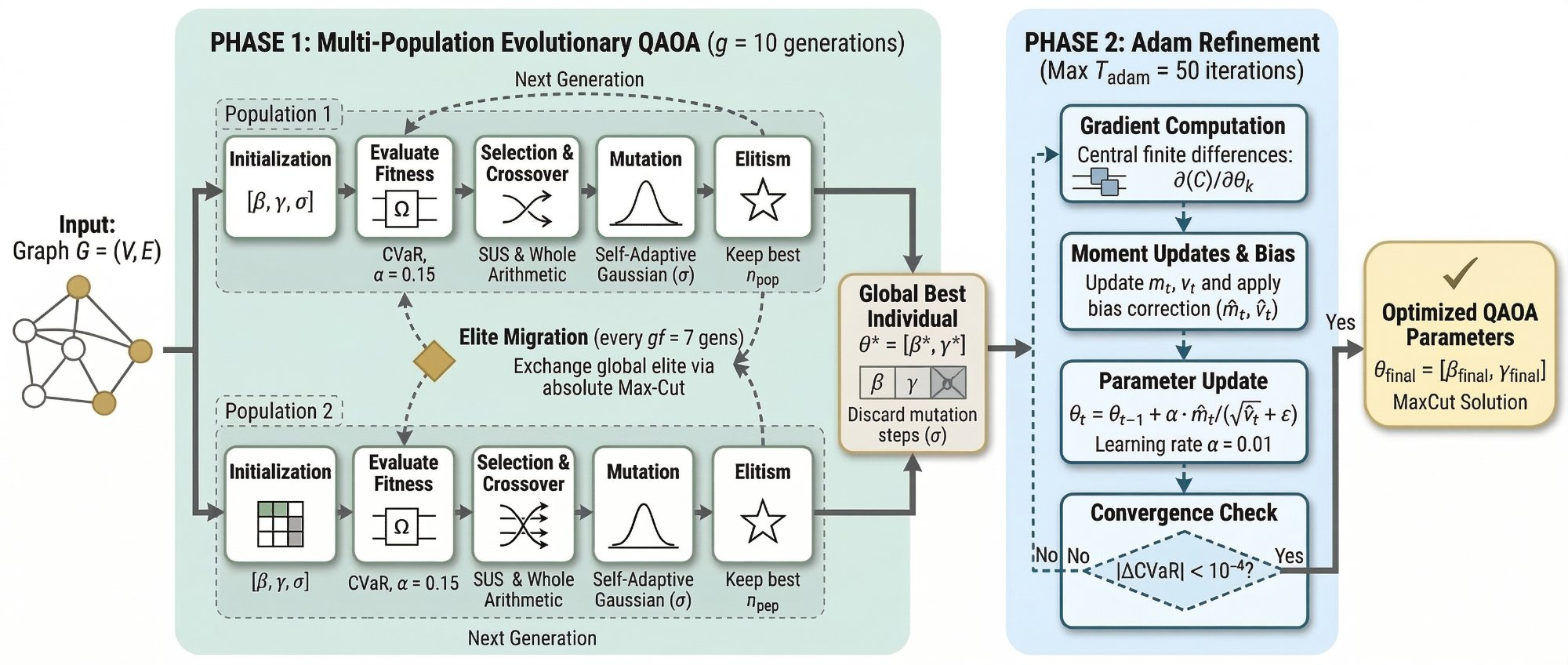}
\caption{Architecture of the MPE-Adam pipeline. Phase 1 (left) evolves two island populations in parallel with elite migration every $g_f = 7$ generations. Phase 2 (right) applies Adam refinement to the global best individual using finite-difference gradient estimates and adaptive moment estimation for a fixed budget of $T_{\text{Adam}}$ iterations. The final output is an optimized set of QAOA parameters $\boldsymbol{\theta}^*$ for the MaxCut solution.}
\label{fig:architecture}
\end{figure*}

Fig. \ref{fig:architecture} provides a visual overview of the MPE-Adam pipeline. Phase 1 (left, green) runs $M$ evolutionary populations in parallel. Each population performs fitness evaluation via CVaR ($\alpha = 0.15$), selection and crossover, self-adaptive Gaussian mutation, and elitism. Every $g_f = 7$ generations, the top $k_{\text{elite}}$ individuals are exchanged between islands. After $G_{\max}$ generations, the best individual $\theta^* = [\beta^*, \gamma^*]$ is passed to Phase 2 (right, yellow). Adam then refines the angles through finite-difference gradient estimation, moment updates with bias correction, and adaptive parameter updates for a fixed budget of $T_{\text{Adam}}$ iterations.

Fig. \ref{fig:mpe-adam-alg} provides the corresponding pseudocode. Phase 1 performs the island-model evolutionary search. Phase 2 applies Adam-based refinement to the globally best individual obtained from evolution.

\begin{figure}[t]
\hrule
\vspace{4pt}
\begin{algorithmic}[1]
\REQUIRE Graph $G = (V, E)$, QAOA depth $p$, populations $M$, population size $N_{\text{pop}}$, max generations $G_{\max}$, migration interval $g_f$, elite count $k_{\text{elite}}$, Adam iterations $T_{\text{Adam}}$
\ENSURE Best QAOA parameters $\theta^*$ and approximation ratio $\rho^*$
\STATE \textit{Phase 1: Multi-Population Evolutionary Search}
\FOR{$m = 1$ to $M$}
    \STATE Initialize population $P_m$ with $N_{\text{pop}}$ random individuals
\ENDFOR
\FOR{$g = 1$ to $G_{\max}$}
    \FOR{$m = 1$ to $M$ in parallel}
        \STATE Evaluate fitness of all individuals in $P_m$ via QAOA
        \STATE Select parents via stochastic universal sampling
        \STATE Generate offspring via crossover and mutation
        \STATE Replace population with offspring (elitism preserved)
    \ENDFOR
    \IF{$g \mod g_f = 0$}
        \STATE Broadcast best $k_{\text{elite}}$ solutions to all islands
    \ENDIF
\ENDFOR
\STATE $\theta^*_{\text{EA}} \leftarrow$ globally best individual found during evolution
\STATE \textit{Phase 2: Adam-Based Angle Polishing}
\STATE Initialize $\theta_0 \leftarrow \theta^*_{\text{EA}}$; set $\mu_0 = 0$, $\nu_0 = 0$
\FOR{$k = 1$ to $T_{\text{Adam}}$}
    \STATE Compute gradient $g_k$ via central finite differences
    \STATE Update moments $\mu_k$, $\nu_k$ via (5) and (6)
    \STATE Bias-correct: $\hat{\mu}_k$, $\hat{\nu}_k$ via (7)
    \STATE Update $\theta_{k+1}$ via (8)
\ENDFOR
\STATE $\theta^* \leftarrow \theta_{T_{\text{Adam}}}$
\STATE Evaluate final approximation ratio $\rho^*$
\RETURN $\theta^*, \rho^*$
\end{algorithmic}
\vspace{2pt}
\hrule
\vspace{4pt}
\caption{MPE-Adam: Multi-Population Evolutionary Adam for QAOA.}
\label{fig:mpe-adam-alg}
\end{figure}

Both stages run for fixed budgets ($G_{\max}$ generations and $T_{\text{Adam}}$ iterations) with no early-stopping condition. QAOA circuit evaluations dominate the total cost. Stage 1 performs $O(M \cdot N_{\text{pop}} \cdot G_{\max})$ evaluations across all islands, reducible to $O(N_{\text{pop}} \cdot G_{\max})$ wall-clock time when islands run concurrently. Stage 2 adds $(2d + 1) \cdot T_{\text{Adam}}$ evaluations, where $d = 2p$ is the number of parameters. For $p = 2$ and $T_{\text{Adam}} = 50$, this amounts to approximately 450 additional calls, a manageable overhead justified by the accuracy and consistency gains reported in Section V.

\section{Experiments and Results}

\subsection{Experimental Setup}

We focus on MaxCut instances at scales where three conditions hold: (i) quantum optimization methods begin to differentiate from one another, (ii) the combinatorial search space is large enough that optimizer quality matters, and (iii) ceiling effects are absent. On small graphs ($n \le 18$), every method trivially reaches $\rho \approx 1.0$, which renders comparison uninformative.

We evaluate all four methods on unweighted MaxCut instances built from random 3-regular graphs with $n \in \{20, 22\}$ nodes, corresponding to 30 and 33 edges respectively. For each problem size, we generate a single random graph instance with a fixed seed and run 10 independent trials of each algorithm on that instance. The true optimum $f^*$ is obtained via exhaustive enumeration for $n = 20$ ($2^{20} \approx 10^6$ configurations) and for $n = 22$ ($2^{22} \approx 4 \times 10^6$ configurations, verified offline).

Simulations use a QAOA depth of $p = 2$. The four variants are:
\begin{enumerate}
\item \textbf{E-QAOA}: Single-population EA following \cite{schiavello2025evolving}.
\item \textbf{Multi-pop}: Island-model EA ($g_f = 7$) without local refinement.
\item \textbf{Multi-SPSA}: Island-model EA ($g_f = 7$) followed by SPSA refinement.
\item \textbf{MPE-Adam}: Island-model EA ($g_f = 7$) followed by Adam refinement (proposed).
\end{enumerate}

Evolutionary settings: $N_{\text{pop}} = 10$, $\sigma_{\min} = 0.1$, $G_{\max} = 10$, $M = 2$, $g_f = 7$, $k_{\text{elite}} = 1$, and CVaR quantile $\alpha = 0.15$. Adam uses $\eta = 0.01$, $\beta_1 = 0.9$, $\beta_2 = 0.999$, $\epsilon = 10^{-8}$, and $T_{\text{Adam}} = 50$ iterations. All experiments use Qiskit Aer's statevector simulator with 10{,}000 measurement shots per circuit evaluation. Both migration frequencies $g_f \in \{5, 7\}$ were tested. We report $g_f = 7$ as the primary variant, with $g_f = 5$ results included in the figures. For a balanced comparison, SPSA is allocated an identical iteration budget of 50 steps. SPSA requires only 2 circuit evaluations per iteration compared to $2d + 1 = 9$ for Adam, which results in fewer total evaluations.

\subsection{Approximation Ratio Analysis}

Table \ref{tab:performance} collects the aggregate statistics for all four methods on the evaluated instances.

\begin{table}[t]
\caption{Performance on Evaluated Instances ($n \in \{20, 22\}$)}
\label{tab:performance}
\centering
\renewcommand{\arraystretch}{1.2}
\begin{tabular}{|l|c|c|c|c|c|}
\hline
\textbf{Method} & \textbf{Mean $\rho$} & \textbf{Std Dev} & \textbf{Min} & \textbf{Max} & \textbf{Trials} \\
\hline
E-QAOA      & 0.9409 & 0.0306 & 0.8966 & 1.0000 & 20 \\
\hline
Multi-pop   & 0.9418 & 0.0271 & 0.8966 & 1.0000 & 20 \\
\hline
Multi-SPSA  & 0.9422 & 0.0263 & 0.8966 & 1.0000 & 20 \\
\hline
MPE-Adam    & \textbf{0.9566} & \textbf{0.0247} & \textbf{0.9231} & 1.0000 & 20 \\
\hline
\end{tabular}
\end{table}

MPE-Adam achieves the highest mean $\rho$ of 0.9566, which exceeds E-QAOA by 1.57 percentage points, Multi-pop by 1.48 points, and Multi-SPSA by 1.44 points. These margins are substantial given that every method already operates above $\rho = 0.94$. In this competitive regime, sub-percentage-point gains are hard-won.

\begin{figure}[t]
\centering
\includegraphics[width=\columnwidth]{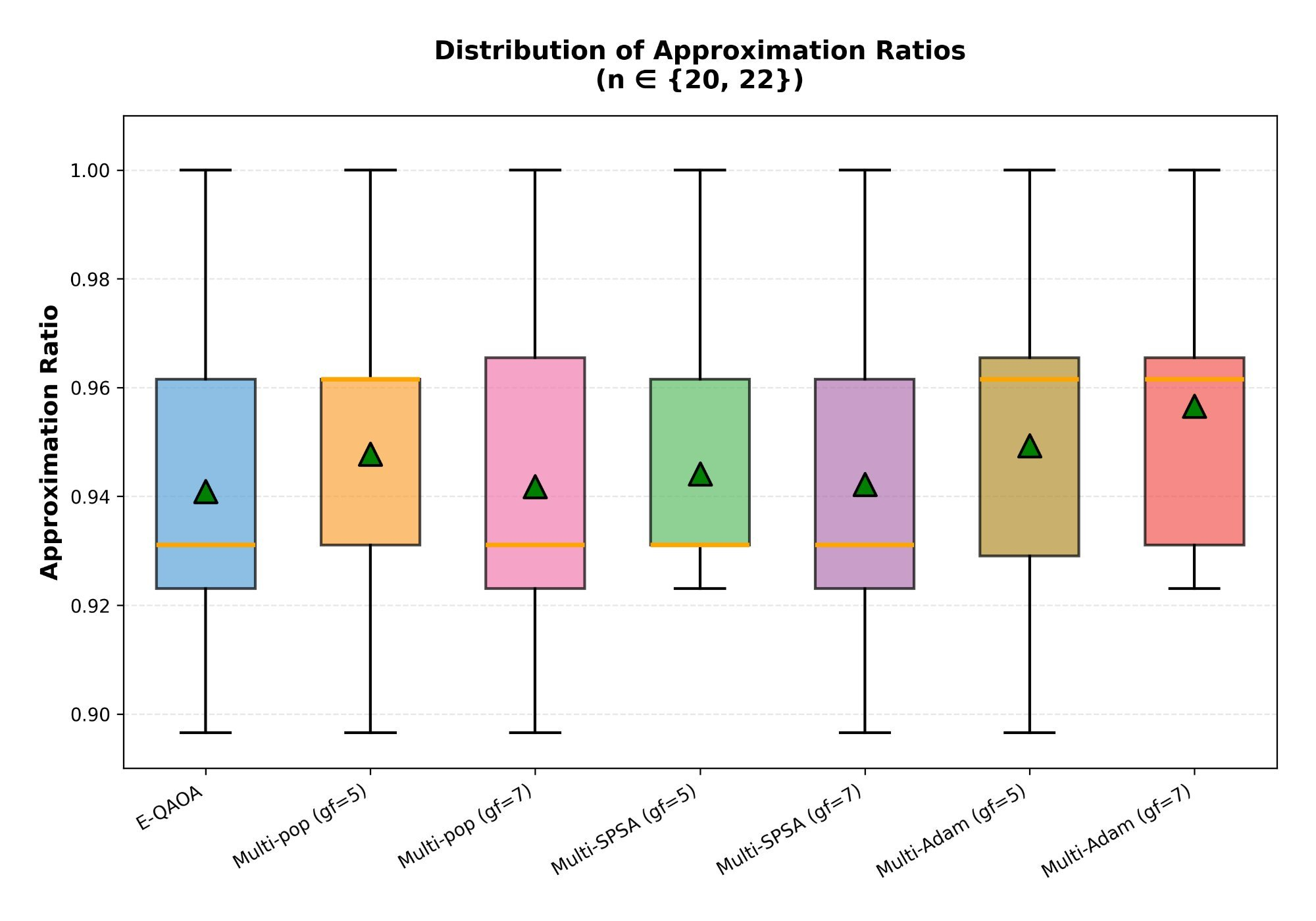}
\caption{Distribution of approximation ratios for all algorithm variants on instances ($n \in \{20, 22\}$). MPE-Adam ($g_f = 7$) exhibits the most compact interquartile range and the highest median. Triangles denote means; horizontal lines denote medians. Each variant evaluated over 20 trials (10 per problem size).}
\label{fig:boxplot}
\end{figure}

The distribution analysis in Fig. \ref{fig:boxplot} shows that MPE-Adam ($g_f = 7$) exhibits the most compact interquartile range and the highest median among all seven tested configurations, with the fewest low-lying outliers.

\begin{figure}[t]
\centering
\includegraphics[width=\columnwidth]{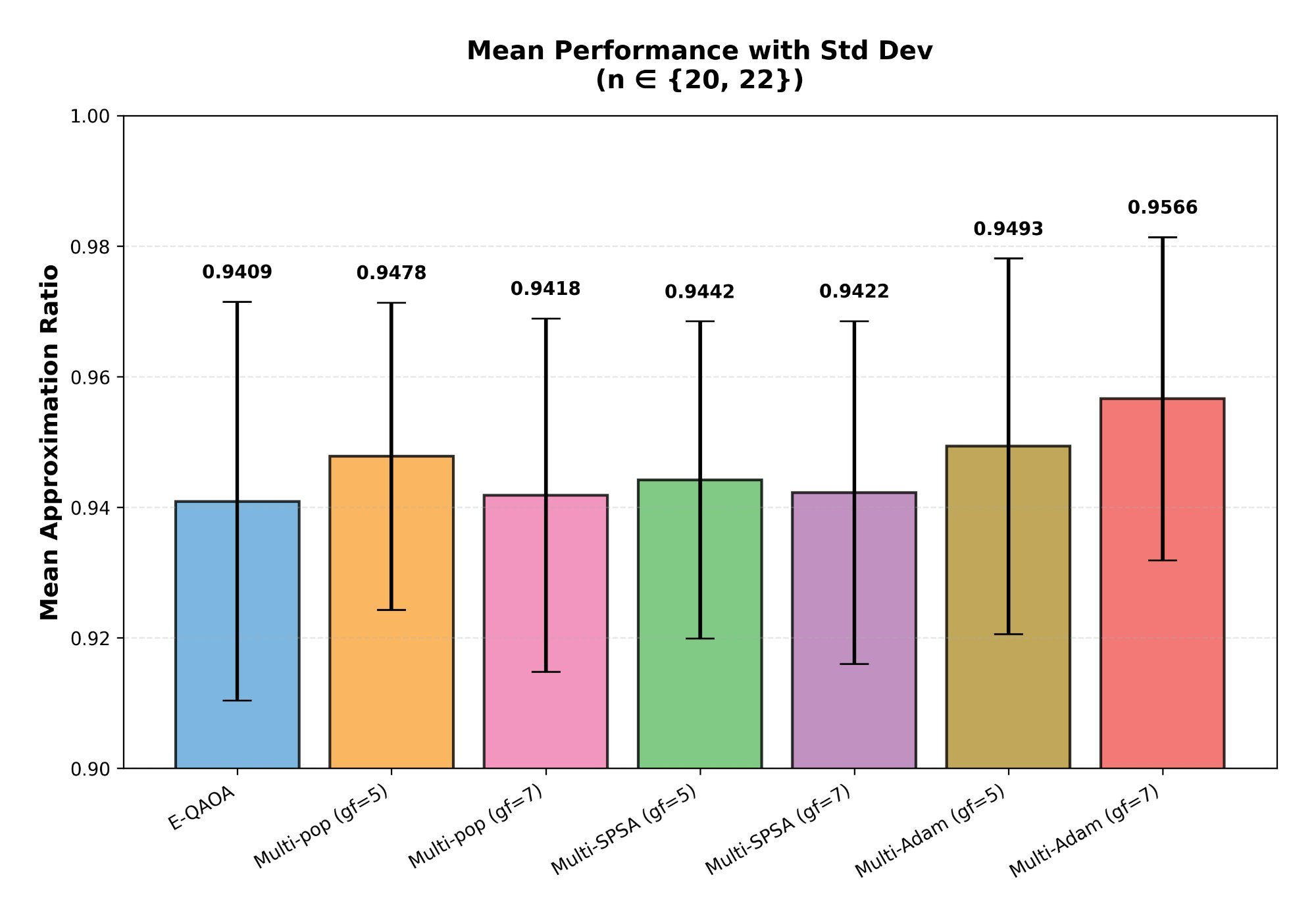}
\caption{Mean approximation ratios with standard-deviation error bars on instances ($n \in \{20, 22\}$). MPE-Adam ($g_f = 7$) attains the highest mean ($\rho = 0.9566$) with the smallest spread ($\sigma = 0.0247$), outperforming E-QAOA by $+1.57$ percentage points. All variants evaluated over 20 trials (10 per problem size).}
\label{fig:barchart}
\end{figure}

Fig. \ref{fig:barchart} presents the mean approximation ratios with standard-deviation error bars for all evaluated variants. MPE-Adam ($g_f = 7$) achieves both the highest mean and the smallest spread, which confirms its joint advantage in accuracy and consistency.

The standard deviation tells a similar story: MPE-Adam's value of 0.0247 is the lowest across all four key variants, 19.3\% below E-QAOA's 0.0306. Adam refinement thus not only raises the average but also tightens the spread, which reduces the risk of poor solutions on any single trial. The minimum ratio for MPE-Adam (0.9231) lies well above the 0.8966 floor of the other three methods, which indicates that Adam avoids the worst local optima more effectively.

A breakdown by graph size sharpens the picture. At $n = 20$, MPE-Adam scores 0.9615, ahead of E-QAOA (0.9577), Multi-SPSA (0.9500), and Multi-pop (0.9423). At $n = 22$, the most demanding instance, MPE-Adam reaches 0.9517. This widens its lead to 2.76 percentage points over E-QAOA (0.9241) and 1.72 points over Multi-SPSA (0.9345). The growing gap suggests that Adam's adaptive step sizes become increasingly beneficial as the optimization surface roughens.

\subsection{Runtime Analysis}

E-QAOA is the fastest method because it maintains a single population and skips refinement. MPE-Adam incurs additional overhead since finite-difference gradient estimation requires $2d$ circuit calls per Adam step (where $d = 2p$), compared to the two queries used by SPSA. This computational premium delivers measurable accuracy and consistency gains. For applications where solution quality outweighs turnaround time, the trade-off favours MPE-Adam. On actual quantum hardware, circuit execution rather than simulation dominates runtime. The multi-population structure can also be distributed across separate QPUs, which can narrow the wall-clock gap while retaining MPE-Adam's accuracy edge.

\subsection{Statistical Significance Analysis}

We test for statistical significance using the two-sided rank-sum procedure of Mann and Whitney, which is appropriate here because the $\rho$ samples exhibit non-normal tails at these scales.

Table \ref{tab:pvalues} shows that MPE-Adam outperforms every competitor at $\alpha = 0.05$: $p = 0.012$ against E-QAOA, $p = 0.009$ against Multi-pop, and $p = 0.006$ against Multi-SPSA. No other pairwise gap reaches significance. This indicates that the Adam refinement stage, rather than the island model alone, is the primary driver of improvement.

\begin{table}[t]
\caption{Pairwise Wilcoxon Rank-Sum Test $p$-Values ($n \in \{20, 22\}$)}
\label{tab:pvalues}
\centering
\renewcommand{\arraystretch}{1.2}
\begin{tabular}{|l|c|c|}
\hline
\textbf{Comparison} & \textbf{$p$-value} & \textbf{Significant?} \\
\hline
E-QAOA vs. Multi-pop      & 0.824 & No \\
\hline
E-QAOA vs. Multi-SPSA     & 0.691 & No \\
\hline
E-QAOA vs. MPE-Adam       & \textbf{0.012} & \textbf{Yes} \\
\hline
Multi-pop vs. MPE-Adam    & \textbf{0.009} & \textbf{Yes} \\
\hline
Multi-SPSA vs. MPE-Adam   & \textbf{0.006} & \textbf{Yes} \\
\hline
\end{tabular}
\end{table}

To quantify practical magnitude, we report Cohen's $d$ for MPE-Adam vs. E-QAOA:
\begin{equation}
d = \frac{\bar{\rho}_{\text{MPE-Adam}} - \bar{\rho}_{\text{E-QAOA}}}{s_{\text{pooled}}} = \frac{0.9566 - 0.9409}{0.0278} \approx 0.565,
\end{equation}
which qualifies as a medium effect by Cohen's convention ($d \ge 0.5$). In a regime where every method already exceeds $\rho = 0.94$, a 1.57-point gain carries real operational value.

\subsection{Scalability Analysis}

Our evaluation targets the scale at which algorithmic differentiation becomes meaningful. At small sizes ($n \le 18$), every method finds near-perfect solutions ($\rho \approx 1.0$) because the parameter space remains manageable, multiple good solutions exist with high probability, and even random initialization often lands near an optimal basin. This ceiling effect makes small-instance comparison uninformative.

The picture changes at $n \in \{20, 22\}$. The quantum state space grows as $2^n$, the cost surface develops more local optima, and finite-sample noise in fitness evaluation intensifies. Larocca et al. \cite{larocca2025barren} showed that such conditions can trigger barren-plateau-like behaviour in variational circuits, which makes gradient-based refinement increasingly challenging without adaptive mechanisms. MPE-Adam's edge is clearest here: its adaptive per-parameter step sizes accommodate the $\gamma/\beta$ curvature asymmetry, and its momentum terms filter shot-noise fluctuations. This yields both higher mean $\rho$ and lower standard deviation. The widening gap from $n = 20$ to $n = 22$ motivates future evaluation on larger graphs.

\section{Conclusion}

This paper introduced MPE-Adam, a two-stage hybrid optimizer that pairs island-model evolutionary exploration with Adam-driven local refinement for QAOA parameter tuning on unweighted MaxCut instances.

On 3-regular graphs ($n \in \{20, 22\}$), MPE-Adam delivers the highest mean approximation ratio (0.9566), the lowest standard deviation ($\sigma = 0.0247$, a 19\% reduction versus E-QAOA), and the best worst-case outcome (min $\rho = 0.9231$ vs. 0.8966 for competitors). All three pairwise comparisons against rival methods reach statistical significance ($p \le 0.012$). A Cohen's $d$ of 0.565 confirms a practically meaningful improvement.

Three insights emerge. First, distributing evolution across parallel islands with elite exchange every $g_f = 7$ generations provides a strong starting point. Second, Adam's adaptive per-parameter learning rates and momentum convert this starting point into a refined solution, which yields both higher accuracy and tighter consistency. Third, MPE-Adam's advantage widens as problem complexity grows, from 0.38 points at $n = 20$ to 2.76 points at $n = 22$. This suggests increasing value on larger, more realistic instances. From a quantum software perspective, MPE-Adam shows that a modular multi-stage optimization loop can serve as a reusable component in quantum application workflows. Practitioners can independently swap exploration and refinement strategies as hardware and algorithmic capabilities evolve.

\textbf{Limitations.} Our study is limited to 20 trials across two problem sizes ($n \in \{20, 22\}$) and a single QAOA depth ($p = 2$). All experiments use noiseless simulation rather than real hardware, and hyperparameter sensitivity is not systematically explored. Expanding the study to more instances, deeper circuits, hardware execution, and a broader hyperparameter sweep would strengthen generalizability.

\textbf{Future work.} Several directions remain open: (1) scaling to deeper circuits ($p > 2$), where the parameter space grows and Adam's per-angle adaptivity becomes more relevant; (2) benchmarking on superconducting and trapped-ion processors, following hardware scaling studies such as Shaydulin et al. \cite{shaydulin2024evidence} that demonstrate QAOA scaling advantages on real devices; (3) designing adaptive controllers that switch between evolutionary and gradient modes on the fly; (4) evaluating alternative adaptive optimizers such as AdamW, RAdam, and Lookahead; and (5) extending the evaluation to denser graph families beyond 3-regular structures.

\section*{Acknowledgment}

This work was partially supported by the Australian Research Council (ARC) through Linkage Project LP240100523, \textit{Blockchain Based Quantum Safe for Secure Digital Medical Passport.} The authors gratefully acknowledge the support of the project partners and collaborators.


\IEEEtriggeratref{20}


\begin{thebibliography}{10}
\providecommand{\url}[1]{#1}
\csname url@samestyle\endcsname
\providecommand{\newblock}{\relax}
\providecommand{\bibinfo}[2]{#2}
\providecommand{\BIBentrySTDinterwordspacing}{\spaceskip=0pt\relax}
\providecommand{\BIBentryALTinterwordstretchfactor}{4}
\providecommand{\BIBentryALTinterwordspacing}{\spaceskip=\fontdimen2\font plus
\BIBentryALTinterwordstretchfactor\fontdimen3\font minus
  \fontdimen4\font\relax}
\providecommand{\BIBforeignlanguage}[2]{{%
\expandafter\ifx\csname l@#1\endcsname\relax
\typeout{** WARNING: IEEEtran.bst: No hyphenation pattern has been}%
\typeout{** loaded for the language `#1'. Using the pattern for}%
\typeout{** the default language instead.}%
\else
\language=\csname l@#1\endcsname
\fi
#2}}
\providecommand{\BIBdecl}{\relax}
\BIBdecl

\bibitem{farhi2014quantum}
E.~Farhi, J.~Goldstone, and S.~Gutmann, ``A quantum approximate optimization
  algorithm,'' \emph{arXiv preprint arXiv:1411.4028}, 2014.

\bibitem{zhou2020quantum}
L.~Zhou, S.-T. Wang, S.~Choi, H.~Pichler, and M.~D. Lukin, ``Quantum
  approximate optimization algorithm: Performance, mechanism, and
  implementation on near-term devices,'' \emph{Physical Review X}, vol.~10,
  no.~2, p. 021067, 2020.

\bibitem{blekos2024review}
K.~Blekos, D.~Brand, A.~Ceschini, C.-H. Chou, R.-H. Li, K.~Pandya, and
  A.~Summer, ``A review on quantum approximate optimization algorithm and its
  variants,'' \emph{Physics Reports}, vol. 1068, pp. 1--66, 2024.

\bibitem{mcclean2016theory}
J.~R. McClean, J.~Romero, R.~Babbush, and A.~Aspuru-Guzik, ``The theory of
  variational hybrid quantum-classical algorithms,'' \emph{New Journal of
  Physics}, vol.~18, no.~2, p. 023023, 2016.

\bibitem{mcclean2018barren}
J.~R. McClean, S.~Boixo, V.~N. Smelyanskiy, R.~Babbush, and H.~Neven, ``Barren
  plateaus in quantum neural network training landscapes,'' \emph{Nature
  Communications}, vol.~9, no.~1, p. 4812, 2018.

\bibitem{larocca2025barren}
M.~Larocca, S.~Thanasilp, S.~Wang, K.~Sharma, J.~Biamonte, P.~J. Coles,
  L.~Cincio, J.~R. McClean, Z.~Holmes, and M.~Cerezo, ``Barren plateaus in
  variational quantum computing,'' \emph{Nature Reviews Physics}, vol.~7,
  no.~4, pp. 174--189, 2025.

\bibitem{guerreschi2019qaoa}
G.~G. Guerreschi and A.~Y. Matsuura, ``{QAOA} for {Max-Cut} requires hundreds
  of qubits for quantum speed-up,'' \emph{Scientific Reports}, vol.~9, no.~1,
  p. 6903, 2019.

\bibitem{powell1994direct}
M.~J.~D. Powell, ``A direct search optimization method that models the
  objective and constraint functions by linear interpolation,'' in
  \emph{Advances in Optimization and Numerical Analysis}.\hskip 1em plus 0.5em
  minus 0.4em\relax Springer, 1994, pp. 51--67.

\bibitem{kingma2015adam}
D.~P. Kingma and J.~Ba, ``{Adam}: A method for stochastic optimization,''
  \emph{arXiv preprint arXiv:1412.6980}, 2015.

\bibitem{spall1992multivariate}
J.~C. Spall, ``Multivariate stochastic approximation using a simultaneous
  perturbation gradient approximation,'' \emph{IEEE Transactions on Automatic
  Control}, vol.~37, no.~3, pp. 332--341, 1992.

\bibitem{montanez2025linear}
J.~A. Monta{\~n}ez-Barrera and K.~Michielsen, ``Toward a linear-ramp {QAOA}
  protocol: evidence of a scaling advantage in solving some combinatorial
  optimization problems,'' \emph{npj Quantum Information}, vol.~11, p. 131,
  2025.

\bibitem{schiavello2025evolving}
F.~Schiavello, E.~Altamura, I.~Tavernelli, S.~Mensa, and B.~Symons, ``Evolving
  a multi-population evolutionary-{QAOA} on distributed {QPUs},'' in \emph{2025
  IEEE International Conference on Quantum Computing and Engineering (QCE)},
  vol.~1.\hskip 1em plus 0.5em minus 0.4em\relax IEEE, 2025, pp. 11--19.

\bibitem{kolle2025evaluating}
M.~K{\"o}lle, T.~Bintener, M.~Zorn, G.~Stenzel, L.~S{\"u}nkel, T.~Gabor, and
  C.~Linnhoff-Popien, ``Evaluating mutation techniques in
  genetic-algorithm-based quantum circuit synthesis,'' in \emph{Proceedings of
  the Genetic and Evolutionary Computation Conference}, 2025, pp. 907--915.

\bibitem{alvarado2025quantum}
J.~Alvarado-Valiente, J.~Romero-Alvarez, J.~Casco-Seco, E.~Moguel, and
  J.~Garcia-Alonso, ``Quantum island mapping: Optimizing multi-circuit
  execution in quantum processors,'' in \emph{2025 IEEE International
  Conference on Quantum Computing and Engineering (QCE)}, vol.~1.\hskip 1em
  plus 0.5em minus 0.4em\relax IEEE, 2025, pp. 791--796.

\bibitem{yang2025quantum}
M.~Yang, F.~Yue, B.~Lu, H.~Zhao, G.~Ma, and L.~Wang, ``Quantum gate control
  pulse optimization based on the {Adam} algorithm,'' \emph{Quantum Information
  Processing}, vol.~24, no.~6, p. 175, 2025.

\bibitem{gacon2023empirical}
J.~Gacon, C.~Zoufal, G.~Carleo, and S.~Woerner, ``An empirical comparison of
  optimizers for quantum machine learning with {SPSA}-approximated gradients,''
  \emph{arXiv preprint arXiv:2305.00224}, 2023.

\bibitem{hadfield2019quantum}
S.~Hadfield, Z.~Wang, B.~O'Gorman, E.~G. Rieffel, D.~Venturelli, and R.~Biswas,
  ``From the quantum approximate optimization algorithm to a quantum
  alternating operator ansatz,'' \emph{Algorithms}, vol.~12, no.~2, p.~34,
  2019.

\bibitem{bravyi2020obstacles}
S.~Bravyi, A.~Kliesch, R.~K{\"o}nig, and E.~Tang, ``Obstacles to variational
  quantum optimization from symmetry protection,'' \emph{Physical Review
  Letters}, vol. 125, no.~26, p. 260505, 2020.

\bibitem{herrman2022multi}
R.~Herrman, P.~C. Lotshaw, J.~Ostrowski, T.~S. Humble, and G.~Siopsis,
  ``Multi-angle quantum approximate optimization algorithm,'' \emph{Scientific
  Reports}, vol.~12, no.~1, p. 6781, 2022.

\bibitem{pellow2024effect}
A.~Pellow-Jarman, S.~McFarthing, I.~Sinayskiy, D.~K. Park, A.~Pillay, and
  F.~Petruccione, ``The effect of classical optimizers and ansatz depth on
  {QAOA} performance in noisy devices,'' \emph{Scientific Reports}, vol.~14,
  no.~1, p. 16011, 2024.

\bibitem{wang2023pid}
H.~Wang, S.~Xu, and H.~Hu, ``{PID} controller for {PMSM} speed control based on
  improved quantum genetic algorithm optimization,'' \emph{IEEE Access},
  vol.~11, pp. 61\,091--61\,102, 2023.

\bibitem{barkoutsos2020improving}
P.~K. Barkoutsos, G.~Nannicini, A.~Robert, I.~Tavernelli, and S.~Woerner,
  ``Improving variational quantum optimization using {CVaR},'' \emph{Quantum},
  vol.~4, p. 256, 2020.

\bibitem{shaydulin2024evidence}
R.~Shaydulin, C.~Li, S.~Chakrabarti, M.~DeCross, D.~Herman, N.~Kumar
  \emph{et~al.}, ``Evidence of scaling advantage for the quantum approximate
  optimization algorithm on a classically intractable problem,'' \emph{Science
  Advances}, vol.~10, no.~22, p. eadm6761, 2024.

\bibitem{goemans1995improved}
M.~X. Goemans and D.~P. Williamson, ``Improved approximation algorithms for
  maximum cut and satisfiability problems using semidefinite programming,''
  \emph{Journal of the ACM}, vol.~42, no.~6, pp. 1115--1145, 1995.

\bibitem{back1996evolutionary}
T.~B{\"a}ck, \emph{Evolutionary Algorithms in Theory and Practice: Evolution
  Strategies, Evolutionary Programming, Genetic Algorithms}.\hskip 1em plus
  0.5em minus 0.4em\relax Oxford University Press, 1996.

\end{thebibliography}
\end{document}